\documentclass[preprintnumbers,amsmath,11pt,amssymb,floatfix,superscriptaddress,nofootinbib]{revtex4}
\usepackage{graphicx}
\usepackage{epsfig}
\usepackage{bm}
\usepackage{amsfonts}

\begin{document}

\title{\Large Generalized second law of thermodynamics in QCD ghost $f(G)$ gravity}

\author{Surajit Chattopadhyay}
\affiliation{ Department of Computer Application, Pailan College
of Management and Technology, Bengal Pailan Park, Kolkata-700 104,
India.} \footnote{$^{,}$$^{*}$ Corresponding author, email:
surajitchatto@outlook.com, surajcha@iucaa.ernet.in}

\begin{abstract}
\textbf{Abstract:} Considering power-law for of scale factor in a flat FRW universe we reported a reconstruction scheme for $f(G)$ gravity based on QCD ghost dark energy. We reconstructed the effective equation of state parameter and observed ``quintessence" behavior of the equation of state parameter. Furthermore, considering dynamical apparent horizon as the enveloping horizon of the universe we have observed that the generalized second law of thermodynamics is valid for this reconstructed $f(G)$ gravity.\\

\end{abstract}


\maketitle

\section{Introduction}
Accelerated expansion of the current universe is well-established by cosmological observations obtained with Supernovae Ia
(SNeIa), the Cosmic Microwave Background (CMB) radiation
anisotropies, the Large Scale Structure (LSS) and X-ray
experiments and is well-documented in literature (Perlmutter et al. 1999; Bennett et al. 2003;
Spergel et al. 2003; Tegmark et al. 2004; Abazajian et al.2004, 2005; Allen et al. 2004). A missing energy component
also dubbed as Dark Energy (DE) characterized by negative pressure
is widely considered by scientists as responsible of this
accelerated expansion. For reviews of DE see Padmanabhan (2005), Copeland et al. (2006), Li et al. (2011) and Bamba et al. (2012). The simplest model of DE is the cosmological constant $\Lambda$ and it is a key
ingredient in the $\Lambda$CDM model. Although the $\Lambda$CDM model is consistent very well with
all observational data, it has the following two weak points as enlisted in Nesseris et al. (2013): (i) It requires a theoretically unnatural and fine-tuned
value for $\Lambda$ and (ii) it is marginally consistent with some recent large
scale cosmological observations. These two weak points of $\Lambda$CDM model have motivated investigation for a wide range of
more complex generalized cosmological models. Different DE candidates have been discussed in Copeland et al. (2006) and Bamba et al. (2012). The issues related to QCD ghost dark energy are discussed in Urban and Zhitnitsky (2010a) and Ohta (2011). The QCD ghost (responsible for the solution of the $UA(1)$ problem) plays a crucial role in the computation of the vacuum energy, because the ghost's properties do not deviate significantly at very large but finite distances (Urban and Zhitnitsky, 2010b).

Importance of modified gravity for late acceleration of
the universe has been reviewed in Nojiri and Odintsov (2007), Tsujikawa (2010) and
Clifton et al. (2012). The theory of scalar-Gauss–Bonnet gravity, named as $f(G)$ has been
proposed by Nojiri and Odintsov (2005). Two noteworthy works on $f(G)$ gravity are Rodrigues et al. (2014) and Houndjo et al. (2013). In the present work we consider a reconstruction scheme for the so-called Gauss-Bonnet gravity, where the gravitational action includes functions of the Gauss-Bonnet invariant $G$ based on QCD ghost dark energy. In this context, we mention the work of Myrzakulov et al. (2011), who studied cosmological solutions, especially the well-known $\Lambda$CDM model in $f(G)$ gravity. In the perspective of studying the accelerated expansion of the current universe, studying reconstruction schemes considering a correspondence between two DE candidates or DE and modified gravity is not new in the literature. Correspondence between holographic dark energy in flat universe and the phantom dark energy model in framework
of Brans–Dicke theory with potential was suggested in Setare (2007a) and Setare (2007b) reconstructed the potential and the dynamics
of the scalar field which describe the Chaplygin cosmology. Another work mentionable in the context of DE reconstruction is Setare (2007c), where a holographic tachyon model was studied in FRW universe. Another noteworthy reference in the area of reconstruction is Setare and Saridakis (2008), where a correspondence between the holographic dark energy scenario in flat universe and the phantom dark energy model in the framework of Gauss-Bonnet
theory with a potential was studied. Jamil and Saridakis (2010) investigated the new agegraphic dark energy scenario in a universe governed by Hořava-Lifshitz gravity. Chattopadhyay and Pasqua (2013) reported a holographic $f(T)$ model and studied its cosmological consequences. Jawad et al. (2013) holographically reconstructed $f(G)$ model and reported quintessence behavior
of effective equation of state parameter. Reconstruction of some cosmological models in $f(R,T)$ gravity was reported in Jamil et al. (2011).

Subsequent sections of the present paper are organized as follows. In section II we have presented the reconstruction methodology for $f(G)$ gravity based on QCD ghost dark energy. In section III we have investigated validity of the generalized second law of thermodynamics for the reconstructed $f(G)$ with dynamical apparent horizon as the enveloping horizon of the universe. We have concluded in section IV.

\section{The reconstruction methodology}
In the work of Sahni and Starobinsky (2006) it was demonstrated that it is possible to rewrite the modified field equations pertaining to modified gravity in the conventional Einsteinian form by transferring all additional terms from the left hand side into the right hand side of the
Einstein equations and referring to them as an effective energy-momentum tensor of dark energy. In the present work we are considering $f(G)$ gravity where the modified field equations are (Sadjadi, 2011; Myrzakulov et al., 2011)
\begin{equation}\label{field1}
3H^2=\rho_G+\rho_m
\end{equation}
\begin{equation}\label{field2}
2\dot{H}+3H^2=p_G+p_m
\end{equation}
where
\begin{equation}\label{rhoGfield}
\rho_G=Gf_G-f(G)-24\dot{G}H^3f_{GG}
\end{equation}
\begin{equation}\label{pGfield}
p_G=-8H^2 \ddot{f}_G-16 H(\dot{H}+H^2)\dot{f}_G-f+Gf_G
\end{equation}
where $f_G=\frac{df}{dG}$ and $f_{GG}=\frac{d^2 f}{dG^2}$ and $G=24(\dot{H}H^2+H^4)$. In view of Sahni and Starobinsky (2006) we observe that Eqs. (\ref{field1}) and (\ref{field2}) along with (\ref{rhoGfield}) and (\ref{pGfield}) are modified field equations, where the additional terms are presented in (\ref{rhoGfield}) and (\ref{pGfield}).

In Eq. (\ref{field1}) we assume that $\rho_G$ behaves like a dynamical dark energy dubbed as ``QCD ghost dark energy" (QCD GDE), whose energy density is proportional to the Hubble parameter (Garcia-Salcedo et al.,2013)
\begin{equation}\label{qde}
\rho_{gde}=\frac{\alpha  (1-\epsilon )}{\tilde{r}_h}=\alpha  (1-\epsilon )\sqrt{H^2+\frac{k}{a^2}};~~\epsilon\equiv\frac{\dot{\tilde{r}}_h}{2H\tilde{r}_h}
\end{equation}
Here, $\alpha$ is a constant with dimension $(energy)^3$ and roughly of order of $\Lambda_{QCD}^3$, where $\Lambda_{QCD}\sim 100 MeV$.  If we ignore the spatial curvature, as we do in this paper, the
trapping horizon is coincident with the Hubble horizon $\tilde{r}_h=1/H$, and
\begin{equation}\label{qdeH}
\rho_{gde}=\alpha  (1-\epsilon ) H;~~\epsilon=-\dot{H}/2H
\end{equation}
Hence, in Eq. (\ref{field1}) we consider $\rho_G=\rho_{gde}$. Our choice for scale factor in the present work is
\begin{equation}\label{a}
a=a_0 t^{n}
\end{equation}
Hence, the Hubble parameter gets the form $H=\frac{n}{t}$. Observing the form of the first Friedmann equation
(\ref{field1}), and comparing to the usual one, we deduce that in
the scenario at hand we obtain an effective dark energy
sector of (modified) gravitational origin. In particular,
one can define the dark energy density and pressure as
\begin{equation}\label{DEdensity}
\rho_{DE}=Gf_G-f(G)-24\dot{G}H^3f_{GG}
\end{equation}
\begin{equation}\label{DEp}
p_{DE}=-8H^2 \ddot{f}_G-16 H(\dot{H}+H^2)\dot{f}_G-f+Gf_G
\end{equation}
In the present work, we are considering DE in the form of QCD GDE. Hence, in (\ref{DEdensity}) we write $\rho_{DE}=\rho_{gde}$ and this leads to a differential equation of the form
\begin{equation}\label{diffe}
4 (1-n)t f+t^2 \left[(6-n) \frac{df}{dt}+t \frac{d^2    f}{dt^2}\right]=4 (n-1)\left(1-\frac{1}{2 n}\right) n \alpha
\end{equation}
Solution of (\ref{diffe}) gives $f(G)$ as a function of $t$ as
\begin{equation}\label{f}
\begin{array}{c}
f(G)=\frac{1}{3nt}\left[3nt^n \left(C_1+C_2 t^{-\frac{3+n}{2}}\right)-2(1-n)(1-2n)\alpha\right]
\end{array}
\end{equation}
Considering $G=\frac{2^{3/4} 3^{1/4} \left((n-1) n^3\right)^{1/4}}{G^{1/4}}$ we can rewrite (\ref{f}) as
\begin{equation}\label{f(G)}
\begin{array}{c}
f(G)=\frac{G^{1/4} }{3\times2^{3/4}
3^{1/4} n \left((n-1) n^3\right)^{1/4}}\left[-2+6 n-4 n^2+2^{3 n/4} 3^{1+\frac{n}{4}} n \left(\frac{\left((n-1) n^3\right)^{1/4}}{G^{1/4}}\right)^n \right.\\ \left.\left(C_1+2^{-\frac{3}{8}
(3+n)} 3^{\frac{1}{8} (-3-n)} C_2 \left(\frac{\left((n-1) n^3\right)^{1/4}}{G^{1/4}}\right)^{\frac{1}{2} (-3-n)}\right)\right]
\end{array}
\end{equation}
Eq. (\ref{f(G)}) is the reconstructed $f(G)$ based on QCD GDE. Subsequent derivatives are
\begin{equation}\label{fG}
\begin{array}{c}
f_G=\frac{1}{48\times
2^{7/8} 3^{5/8} G^{3/4} n \left((n-1) n^3\right)^{3/4}}
\left[-2^{3 n/8} 3^{1+\frac{n}{8}}C_2 \sqrt{G} (-5+n) n \left(\frac{\left((n-1) n^3\right)^{1/4}}{G^{1/4}}\right)^{\frac{1+n}{2}}\right.\\
\left.-2^{17/8} 3^{3/8} (n-1) \sqrt{(n-1) n^3} \left\{-2+n \left(4+2^{3 n/4} 3^{1+\frac{n}{4}}C_1 \left(\frac{\left((n-1) n^3\right)^{1/4}}{G^{1/4}}\right)^n\right)\right\}\right]
\end{array}
\end{equation}
\begin{equation}\label{fGG}
\begin{array}{c}
f_{GG}=\frac{1}{128\times 2^{7/8} 3^{5/8} G^{7/4} n \left((n-1) n^3\right)^{3/4}}\left[2^{3 n/8} 3^{n/8} C_2 \sqrt{G} (-5+n) n \left(\frac{\left((n-1) n^3\right)^{1/4}}{G^{1/4}}\right)^{\frac{1+n}{2}}\right.\\
\left.(3+n)+8 2^{1/8} 3^{3/8} (n-1) \sqrt{(n-1) n^3} \left(-2+4 n+2^{3 n/4} 3^{n/4} C_1 n \left(\frac{\left((n-1) n^3\right)^{1/4}}{G^{1/4}}\right)^n
(3+n)\right)\right]
\end{array}
\end{equation}

\begin{figure}[h]\label{fplot}
\includegraphics[width=28pc]{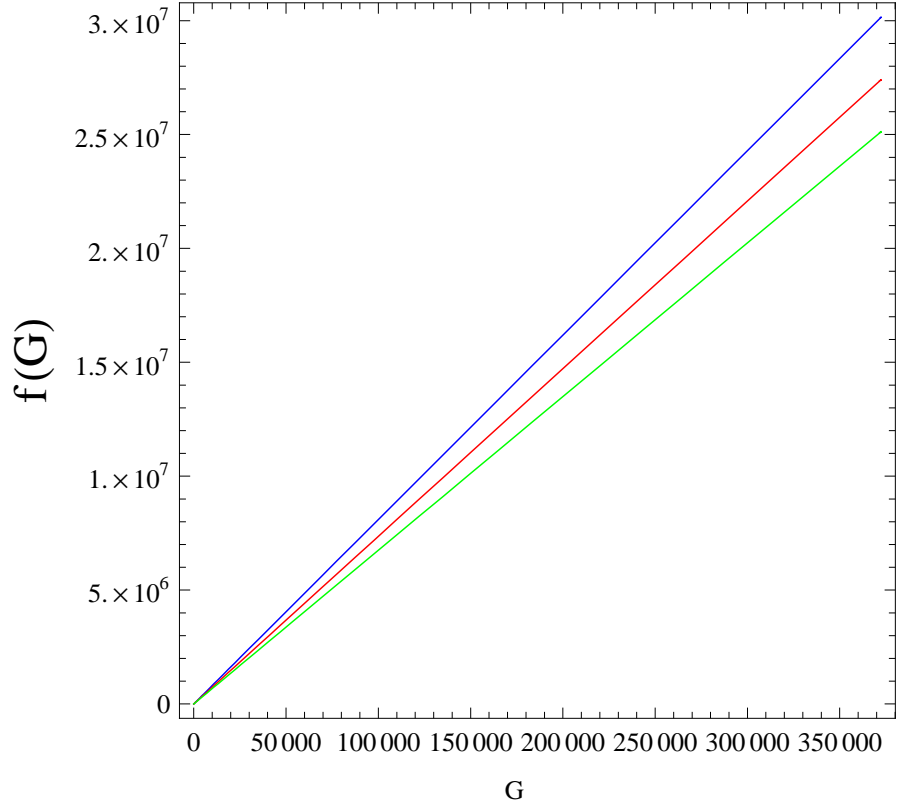}
\caption{\label{label}Plot of $f(G)$ against $G$ for some values of $n>1$ and it is observed that $f(G)\rightarrow 0$ as $G\rightarrow 0$. }
\end{figure}
The reconstructed $f(G)$ presented in Eq. (\ref{f(G)}) is plotted against $G$ in Fig. \ref{fplot}, where we have taken $n=1.01,~C_1=-10.02,~C_2=20.02,~\rho_{m0}=0.23$ and $a_0=12.1$. The red, green and blue lines correspond to $\alpha=1.1,~1.5$ and $1.03$ respectively. It is observed in the figure that $f(G)\rightarrow 0$ as $G\rightarrow 0$. This indicates that the reconstructed $f(G)$ represents a realistic model.

Using Eqs. (\ref{f(G)}), (\ref{fG}) and (\ref{fGG}) in (\ref{rhoGfield}) we get reconstructed $\rho_G$ as
\begin{equation}\label{rhoG}
\rho_G=\frac{G^{1/4} (2 n-1) \alpha }{2^{7/4} 3^{1/4} \left\{(n-1) n^3\right\}^{1/4}}
\end{equation}
In Eq. (\ref{rhoG}) it is essential that $n>1$. Similarly, using Eqs. (\ref{f(G)}), (\ref{fG}) and (\ref{fGG}) in (\ref{pGfield}) we get reconstructed $p_G$ as
\begin{equation}\label{pG}
p_G=-\frac{G^{1/4} (2n-1)(3n-1) \alpha }{2^{7/4} 3^{5/4} n \left\{(n-1) n^3\right\}^{1/4}}
\end{equation}
In Eq. (\ref{pG}), $n>1\Rightarrow p_G<0$.
Conservation equation for pressureless ($p_m=0$) dark matter is
\begin{equation}\label{conservation}
\dot{\rho}_m+3H(\rho_m)=0
\end{equation}
that gives
\begin{equation}\label{rhom}
\rho_m=\rho_{m0}a^{-3}
\end{equation}
Using Eqs. (\ref{rhoG}), (\ref{pG}) and (\ref{rhom}) the effective EoS parameter is found to be
\begin{equation}\label{w}
w_{eff}=\frac{p_G}{\rho_G+\rho_m}=\frac{(1-2n)(1-3n)}{n} \left[3-6 n-\frac{2^{\frac{7-9n}{4}} 3^{\frac{5-3n}{4}} \left(\frac{\left\{(n-1) n^3\right\}^{1/4}}{G^{1/4}}\right)^{1-3
n} \rho_{m0}}{a_0^3 \alpha }\right]^{-1}
\end{equation}
\begin{figure}[h]\label{plotw}
\includegraphics[width=28pc]{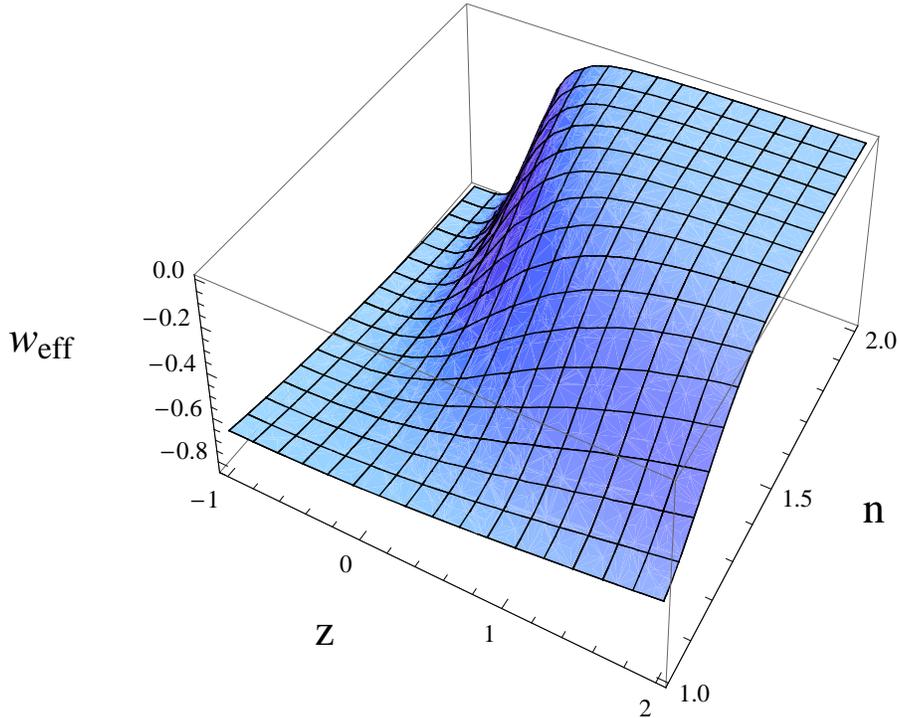}
\caption{\label{eosplot}Evolution of effective EoS parameter $w_{eff}$ against redshift
$z=a^{-1}-1$ and a range of values of $n>1$ based on Eq. (\ref{w}). }
\end{figure}
The EoS parameter presented in Eq. (\ref{w}) is plotted in Fig. \ref{plotw}, where we have taken $\alpha=1.3,~C_1=-10.02,~C_2=20.02,~\rho_{m0}=0.23$ and $a_0=12.1$. We have presented $w_{eff}$ in a 3D plot for a range of values on $n>1$ over redshift $z=a^{-1}-1$. It is observed that for smaller values of $n$, $w_{eff}>-1$ over the entire range of $z$. However, for larger values of $n$ e.g. $n\approx 2$, $w_{eff}\rightarrow -1$ with evolution of the universe i.e. reaches phantom boundary, but never crosses it. Currently, i.e. for $z=0$, $w_{eff}>-1~\forall~n>1$. Thus, in general, the $w_{eff}$ behaves like quintessence (Cai et al., 2010).
\section{GSL in QCD ghost $f(G)$ gravity}
Since the discovery of black hole thermodynamics in 1970s, it has been speculated by physicists that there should be some relationship between thermodynamics and Einstein equations because the horizon area (geometric quantity) of black hole is associated with its entropy (thermodynamical quantity), the surface gravity (geometric quantity) is related with its temperature (thermodynamical quantity) in black hole thermodynamics (Akbar and Cai, 2006; Jamil et al., 2012). In analogy with black hole thermodynamics, investigating thermodynamics laws for cosmological horizons, which are present in many of cosmological
models, has also been the subject of many studies (e.g. Busso, 2005; Jamil et al., 2010 (a, b); Sadjadi, 2011; Saridakis et al., 2009; Debnath et al., 2012; Jamil et al., 2012a). Here we mention some examples of exploration of the generalized second law (GSL) of thermodynamics in modified gravity scenario. Bamba and Geng (2009) demonstrated that an $f(R)$ gravity model with realizing a crossing of the phantom divide can satisfy the GSL in the effective phantom phase as well as non-phantom one. Bamba and Geng (2011) explored thermodynamics of the apparent horizon in $f(T)$ gravity with both equilibrium and non-equilibrium descriptions.  Karami and Abdolmaleki (2012)  investigated the validity of the generalized second law (GSL) of gravitational thermodynamics in the framework of $f(T)$ modified teleparallel gravity. Chattopadhyay and Ghosh (2012) established validity of GSL in the modified f(R) Horava–Lifshitz gravity.

Here, we investigate the validity of the GSL for a spatially flat FRW universe. Hawking temperature on the apparent horizon $\tilde{r}_A$ is given by
\begin{equation}
T_A=\frac{n}{2 \pi  t}\left(1-\frac{1}{2 n}\right)
\end{equation}
where $n>1/2$ ensures $T_A>0$.

The entropy of the universe including the dark matter inside the dynamical apparent horizon is given by Gibb's equation
\begin{equation}\label{gibbs}
T_A dS_m=dE_m+p_m dV
\end{equation}
where, $p_m=0$ and $V=4\pi \tilde{r}_A^3/3$ is the volume containing the pressureless dust matter with the radius of the dynamical apparent horizon $\tilde{r}_A$. Also,
\begin{equation}\label{Em}
E_m=\frac{4\pi \tilde{r}_A^3}{3}\rho_m
\end{equation}
From the modified field equation if we write $\rho_m =3 H^2-\rho_G$ then we can have
\begin{equation}
\begin{array}{c}
T_A \dot{S}_m=4\pi \tilde{r}_{A}^2 \rho_{m}(\dot{\tilde{r}}_{A}-H \tilde{r}_{A})=-\frac{2 (n-1) \pi  }{n^3}\left[6 n^2+t \alpha (1-2 n) \right]
\end{array}
\end{equation}
In the thermodynamics of the apparent horizon in the Einstein gravity, the geometric
entropy is assumed to be proportional to its horizon area $S_A=2 \pi A$. However, this
definition is changed for other modified gravity theories. The geometric entropy in
$f(R)$ gravity is given by $S_A=2 \pi A f_R$, where $f_R=\frac{df}{dR}$. In $f(T)$ gravity, it was shown that when $f_{TT}$ is small, the first law of
black hole thermodynamics is satisfied approximatively and the entropy of horizon is $S_A=2 \pi A f_T$, where $f_T=\frac{df}{dT}$. In the present case, where we are considering $f(G)$ gravity, we compute
\begin{equation}\label{SA}
S_A=2 \pi A f_G=-\frac{\pi ^2 t^2 \left(-12 C_2 n+(n-1) t^3 \left(3 C_1 n t^n-2 \alpha +4 n \alpha \right)\right)}{36 (n-1) n^6}
\end{equation}
From (\ref{SA}) it is easy to get
\begin{equation}
\begin{array}{c}
T_A \dot{S}_A=\frac{(2 n-1) \pi }{144 (1-n) n^6}\left[-24 C_2 n+(n-1) t^3 \left(3 C_1 n (5+n) t^n+10 (-1+2 n) \alpha \right)\right]
\end{array}
\end{equation}
\begin{equation}
\begin{array}{c}\label{gsl}
\dot{S}_{total}=\dot{S}_{A}+\dot{S}_{m}=\\
-\frac{4 \pi^2  t}{2n-1} \left[\frac{2 (n-1) \left(6 n^2+t \alpha -2 n t \alpha \right)}{n^3}+\frac{(-1+2 n)  \left(-24 C_2 n+(n-1)
t^3 \left(3 C_1 n (5+n) t^n+10 (-1+2 n) \alpha \right)\right)}{144 (n-1) n^6}\right]
\end{array}
\end{equation}

\begin{figure}[h]
\includegraphics[width=28pc]{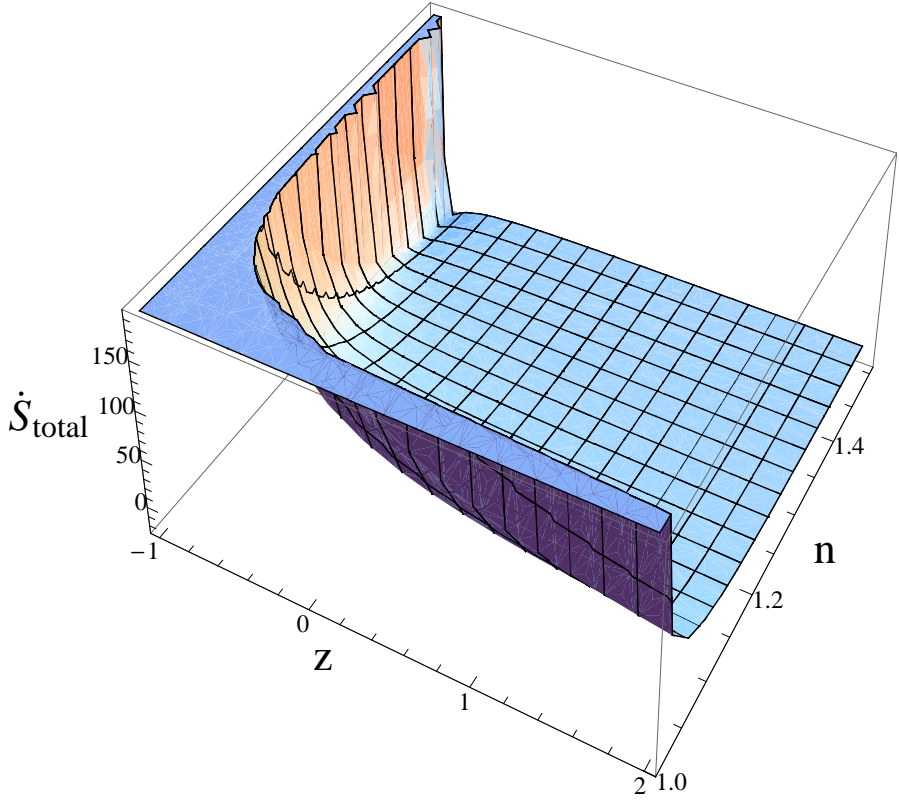}
\caption{\label{sdotplot}Evolution of time derivative of total entropy $\dot{S}_{total}$ based on (\ref{gsl}) against
redshift $z=a^{-1}-1$ and a range of values of $n>1$.}
\end{figure}
The time derivative of total entropy $\dot{S}_{total}$  derived in Eq. (\ref{gsl}) for the QCD ghost $f(G)$ gravity given by Eq. (\ref{f(G)}) is presented in   Fig. \ref{sdotplot} for a range of $n>1$ against redshift $z$ and it is observed that for the entire range of $n$ we have $\dot{S}_{total}>0$ and this indicates validity of GSL for this QCD ghost $f(G)$ gravity with apparent horizon as the enveloping horizon of the universe. It is also observed that $\dot{S}_{total}$ is decreasing with $n$ for a given redshift and $\dot{S}_{total}$ is increasing with evolution of the universe.

\section{Concluding remarks}
In the present study we have presented a reconstruction scheme for $f(G)$ gravity, where $G$ represents Gauss-Bonnet invariant and we have chosen the scale factor in power-law form. Considering a flat FRW universe, we have considered a correspondence between $f(G)$ gravity and QCD ghost dark energy. After getting a reconstructed solution for $f(G)$ in Eq, (\ref{f(G)}) we have plotted $f(G)$ against $G$, where it is apparent that $f(G)\rightarrow~0$ as $G\rightarrow~0$, which is one of the sufficient conditions for a realistic model. In Eq. (\ref{w}) we have plotted the effective equation of state parameter for $n>1$ in Fig. \ref{eosplot} and observed that $w_{eff}\geq~-1$ i.e. it behaves like quintessence. For $n\approx~2$, $w_{eff}$ is reaching $-1$ at late stage of the universe. However, it is not crossing the phantom boundary. Subsequently, considering dynamical apparent horizon as the enveloping horizon of the universe we have derived expression for the time derivative of total entropy $\dot{S}_{total}$ in Eq. (\ref{gsl}) and we have observed that $\dot{S}_{total}>0$ throughout the evolution of the universe. This indicated validity of the generalized second law of thermodynamics under this reconstruction of $f(G)$ gravity.
\subsection{Acknowledgement}
Financial support from Department of Science and Technology, Govt. of India, under Project Grant No. SR/FTP/PS-167/2011 is duly acknowledged.

\begin{flushleft}
\textbf{References}\\
\end{flushleft}
Akbar, A., Cai, R. G.: Phys. Lett. B 635, 7 (2006)\\
Abazajian, K., et al.: Astron. J. 128, 502 (2004)\\
Abazajian, K., et al.: Astron. J. 129, 1755 (2005)\\
Allen, S.W., Schmidt, R.W., Ebeling, H., Fabian, A.C., van Speybroeck,
L.: Mon. Not. R. Astron. Soc. 353, 457 (2004)\\
Bamba, K., Capozziello, S., Nojiri, S., Odintsov, S. D.: Astrophys. Space Sci. 341, 155 (2012)\\
Bamba, K., Geng, C-Q.: JCAP 11, 008 (2011) doi:10.1088/1475-7516/2011/11/008\\
Bamba, K., Geng, C-Q.:  Phys. Lett. B 679, 282 (2009)\\
Bennett, C.L., et al.: Astrophys. J. 583, 1 (2003)\\
Bousso, R.: Phys. Rev. D 71, 064024 (2005)\\
Cai, Y. F., Saridakis, E. N., Setare, M. R., Xia, J. Q.: Physics Reports, 493, 1 (2010)\\
Copeland, E.J., Sami, M., Tsujikawa. S.: Int. J. Mod. Phys. D 15, 1753 (2006)\\
Chattopadhyay, S., Pasqua, A.: Astrophys. Space Sci. 344, 269 (2013)\\
Chattopadhyay, S.,  Ghosh, R.: Astrophys. Space Sci. 341, 669 (2012)\\
Clifton, T., Ferreira, P.G., Padilla, A., Skordis, C.: Phys. Rep. 513, 1
(2012)\\
Debnath, U., Chattopadhyay, S., Hussain, I., Jamil, M., Myrzakulov, R.: Eur. Phys. J. C  72, 1 (2012) 10.1140/epjc/s10052-012-1875-7\\
Garcia-Salcedo, R., Gonzalez, T., Quiros, I., Thompson-Montero, M.: Phys. Rev. D 88, 043008 (2013)\\
Houndjo, M. J. S., Rodrigues, M. E., Momeni, D.,  Myrzakulov, R. arXiv preprint arXiv:1301.4642 (2013)\\
Jamil,  M., Momeni, D., Myrzakulov, R.:   Eur. Phys. J. C  72,  2137 (2012)[arXiv:1210.0001 [physics.gen-ph]]\\
Jamil, M., Yesmakhanova, K., Momeni, D., Myrzakulov, R.: Central Eur. J. Phys.  10, 1065 (2012)
  [arXiv:1207.2735 [gr-qc]].\\
Jamil, M., Saridakis, E. N., Setare, M. R.: Phys. Rev. D 81, 023007 (2010a)\\
Jamil, M., Saridakis, E. N., Setare, M. R.: JCAP 11, 032 (2010b) doi:10.1088/1475-7516/2010/11/032\\
Jamil, M., Momeni, D.,  Myrzakulov, R. Chinese Physics Letters 29, 109801 (2012)\\
Jawad, A., Pasqua, A., Chattopadhyay, S.: Astrophys. Space Sci. 344, 489 (2013)\\
Jamil, M., Momeni, D., Raza, M.,  Myrzakulov, R.: Eur. Phys. J. C 72, 1 (2012a)\\
Jamil, M., Momeni, D., Bamba, K., Myrzakulov, R. :  Int. J. Mod. Phys. D 21, 1250065 (2012) DOI: 10.1142/S0218271812500654 \\
Li, M., et al.: Commun. Theor. Phys. 56, 525 (2011)\\
Myrzakulov, R., Sáez-Gómez, D., Tureanu, A.: Gen. Relativ. Gravit.
43, 1671 (2011)\\
Nesseris, S., Basilakos, S., Saridakis, E. N., Perivolaropoulos, L.: Phys. Rev. D 88, 103010 (2013)\\
Nojiri, S., Odintsov, S.D.: Int. J. Geom. Methods Mod. Phys. 4, 115
(2007)\\
Nojiri, S., Odintsov, S.D.: Phys. Lett. B 631, 1 (2005)\\
Ohta, N.: Phys. Lett. B, 695, 41 (2011)\\
Padmanabhan, T.: Curr. Sci. 88, 1057 (2005)\\
Perlmutter, S., et al.: Astrophys. J. 517, 565 (1999)\\
Rodrigues, M.~E.~, Houndjo, M.~J.~S.~,Momeni, D., Myrzakulov, R.: Can. J. Phys. 92, 173 (2014)\\
Sadjadi, H. M.,  Phys. Scr. 83, 055006 (2011)\\
Saridakis, E. N., Gonzalez-Diaz, P. F.,  Sigüenza, C. L.: Class. Quantum Grav. 26, 165003 (2009) doi:10.1088/0264-9381/26/16/165003\\
Setare, M.R.: Phys. Lett. B 644, 99 (2007a)\\
Setare, M.R.: Phys. Lett. B 648, 329 (2007b)\\
Setare, M.R.: Phys. Lett. B 653, 116 (2007c)\\
Setare, M.R. and Saridakis, E. N.: Phys. Lett. B 670, 1 (2008)\\
Spergel, D.N., et al.: Astrophys. J. Suppl. Ser. 148, 175 (2003)\\
’t Hooft, G.: arXiv:gr-qc/9310026 (1993)\\
Tegmark, M., et al.: Phys. Rev. D 69, 103501 (2004)\\
Urban, F. R., Zhitnitsky, A. R.: Nucl. Phys. B, 835, 135 (2010a)\\
Urban, F. R., Zhitnitsky, A. R.: Phys. Lett. B, 688, 9 (2010b)\\

\end{document}